\documentclass{appolb}
\usepackage{epsfig}

\begin{document}
\title{ON THE NEW HERMES DATA \\
FOR THE ELECTROPRODUCTION ON NUCLEI%
}
\author{K. Fia{\l}kowski and R. Wit
\address{M. Smoluchowski Institute of Physics, Jagellonian University \\
30-059 Krak{\'o}w, ul. Reymonta 4, Poland} }
\maketitle
\begin{abstract}
We analyze recent data on the electroproduction of hadrons on
nuclei using the Lund model for electroproduction on nucleons and
a simple geometrical model for the absorption effects. We show
that the model seems to overestimate the $A$-dependence of the
absorption effects, although it described the earlier data of the
same HERMES experiment reasonably well. We trace the origin of
this discrepancy to the surprising difference between the data for
nitrogen and neon.
\end{abstract}

\section{Introduction}
   ~~~~In a recent paper \cite{FW} we presented a comparison of the
   data from the HERMES collaboration on the
electroproduction on $N$, $Kr$ and $Xe$ nuclei both for the single
spectra \cite{H1} and for the two hadron systems \cite{H2} with a
simple model based on the PYTHIA \cite{PYTHIA} code  for the
electroproduction on nucleons and the geometrical scheme for
calculating the absorption effects. We investigated the
ratios of spectra for which many systematic uncertainties cancel.
\par
 We discussed a very simple
picture, in which only the obvious part of the Lund space-time
development is used, and we supplemented it with (equally obvious)
pure absorptive effects. We restricted
ourselves to the use of hadronic (and not partonic) degrees of
freedom, since we discussed the low energy data for which the
typical $Q^2$ values are small.
\par
Surprisingly, we found a reasonably good description of data for
the ratios of single spectra of charged hadrons. With only one
free parameter, a "hadronization proper time" $\tau_h$, the
dependence on the relative energy $z=E_h/\nu$ is well described
both for nitrogen ($A=14$) and krypton ($A=84$) for the range of
$z$ in which the non-absorptive effects may be neglected ($z>0.1$
and $z>0.3$, respectively). These data are dominated by pions, but
for identified kaons the krypton data were also compatible with
model predictions using smaller value of $\tau_h$ (as expected for
heavier particles). Even the data for "second fastest" hadron are
qualitatively compatible with the model for $Kr$ and $Xe$ nuclei
in similar range of $z$. We did not try to compare the model with
the data as functions of $Q^2$ or $\nu$, since it is rather
difficult to estimate the limits of applicability of a purely
absorptive model for these variables.
\par
    Recently the HERMES collaboration presented a new version of data
\cite{H3} with the identification of pions, kaons and
(anti)protons in the full range of $z>0.1$ for $He,$ $Ne,$ $Kr$
and $Xe$ nuclei. In the next section we present a comparison of
these data with the predictions of our model (with no new
parameters). The conclusions are included in the last section.

\section{The model and the data}

~~~~As before, we are using the Monte Carlo generator PYTHIA 6.203
 and generate more than a quarter million of events
per each nucleus, applying all the kinematical cuts from HERMES
data, either by setting the proper values of PYTHIA
parameters, or explicitly in the program for the event analysis.
\par
We supplement the ordinary information provided by PYTHIA for each
event by extracting the values of one extra parameter from the generating
algorithm: the GAM(3) parameter, set for $each$ string break in
the PYSTRF procedure and denoting the proper time $\tau_0$ (time
measured in the string rest frame) between the string formation
and its break. This time, corrected for the Lorentz dilatation, is
used to calculate the distance between the string formation and
string breaking point in the nucleus rest frame
$$s^0_{form}=\tau_0v_{str}\gamma_{str}.$$
To account for the time needed to rearrange partons from the break
into hadrons, we introduce the only free parameter of our model, a
"hadronization proper time" $\tau_h$ (found to be $0.7-0.8$ fm for
pions, and $0.3-0.4$ fm for kaons)), which is subsequently
dilatated by a $string$ Lorentz factor $\gamma_h$

$$s_{form}=(\tau_0+\tau_h)v_{str}\gamma_{str}.$$

\par
The generation of the string creation point inside nucleus and the
calculation of the absorption factor is performed as described in
our previous paper \cite{FW}.

\begin{figure}[h]
\centerline{%
 \epsfig{figure=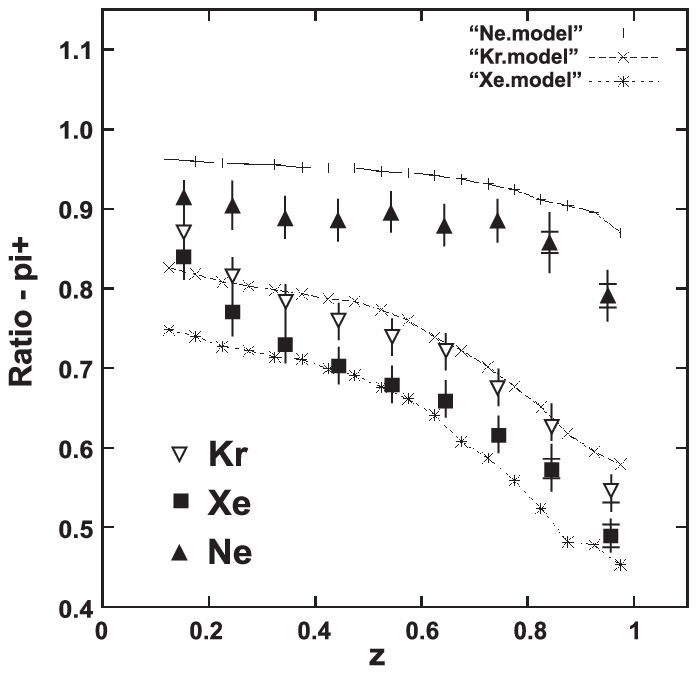, height =7cm}
\hspace{1cm}
 \epsfig{figure=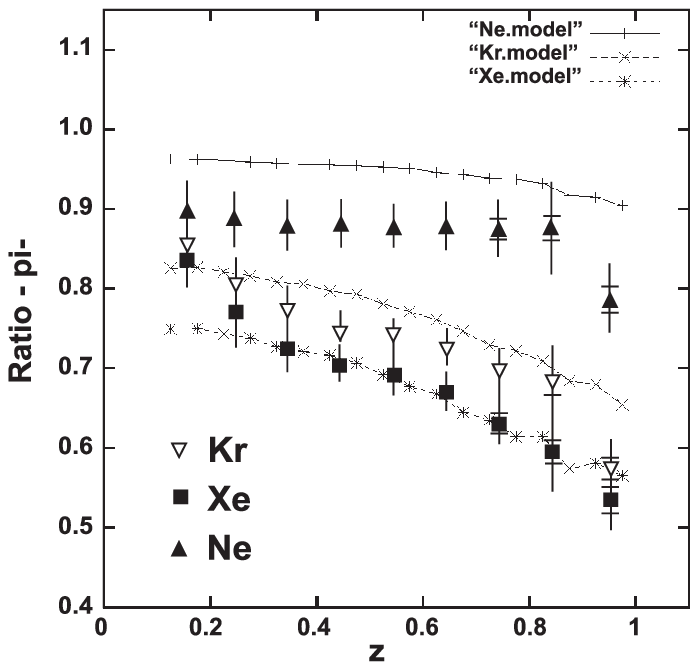,height=7cm}}
 \caption{\label{fig1} {\small \sl The experimental ratio of the
$\pi^+$ (on the left) and $\pi^-$ (on the right)  $z$-spectra from
neon, krypton and xenon to that from the deuterium \cite{H3}
compared with the model calculations for $\tau_h=0.8$ fm/c.}}
\end{figure}

\begin{figure}[h]
\centerline{%
\epsfig{figure=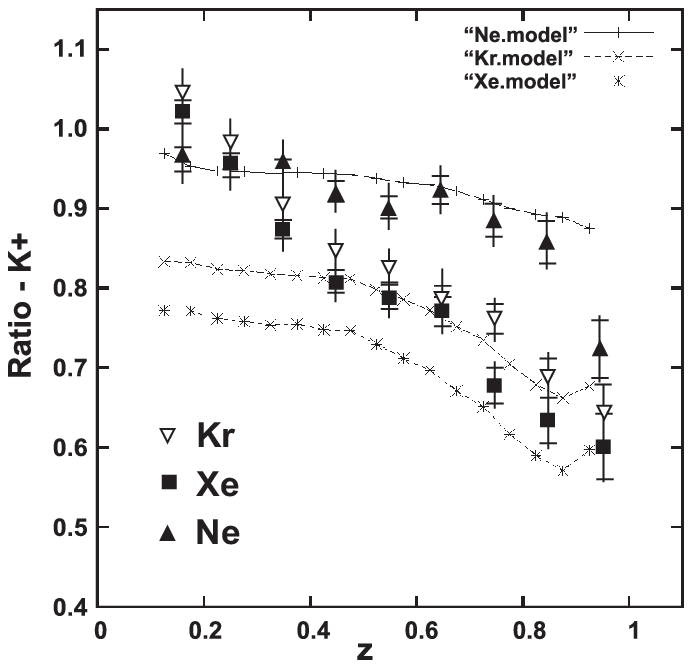, height =7cm} \hspace{1cm}
\epsfig{figure=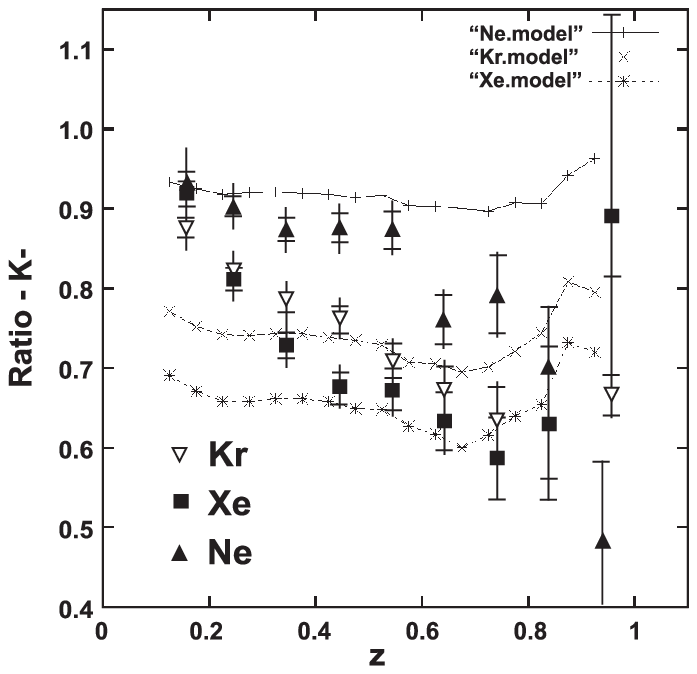, height =7cm}} \caption{\label{fig2} {\small
\sl The experimental ratio of the $K^+$ (on the left) and $K^-$
(on the right)  $z$-spectra from neon, krypton and xenon to that
from the deuterium \cite{H3} compared with the model calculations
for $\tau_h=0.4$ fm/c.}}
\end{figure}

\par
In Figs. 1. and 2. we show the ratios of the properly normalized
single spectra

$$ R^h(z,\nu,p_t^2,Q^2)=\Big(\frac{N_h(z,\nu,p_t^2,Q^2)}{N_e(\nu,Q^2)}\Big)_A\Big/
\Big(\frac{N_h(z,\nu,p_t^2,Q^2)}{N_e(\nu,Q^2)}\Big)_d$$

\noindent for pions and kaons produced on $Ne,$ $Kr$ and $Xe$. The
$He$ data, which are compatible within errors with no significant
absorption effects both in the data and in the model, are omitted
for transparency. We do not show the data for protons, as our
purely absorptive model is obviously unable to reproduce them.

\par
We see that the agreement of the model with data for the $Ne$
nucleus is  poor: the data  are significantly below the
predictions. In addition, the data for pions (which dominate the
spectra) show much weaker dependence on the atomic mass $A$ than
expected from the model. The data for xenon are  above the
predictions. The change of the value of the only free parameter of
the model, $\tau_h$, cannot improve the situation: e.g. for
$\tau_h=0.6$ fm the model agrees with neon data, but overestimates
the absorption effects for both heavier nuclei . For kaons the
agreement is even worse and a similar pattern is seen. In both
cases the model is not applicable for $Kr$ and $Xe$ when $z \leq
0.3$ since no secondary production is included.

\par
This disagreement is surprising in view of the successes of the
model for the previous data. Thus we decided to compare the data
and model predictions for two light nuclei: nitrogen \cite{H1} and
neon \cite{H2}. For nitrogen all charged hadrons are counted; for
neon positive pion spectra are shown. This is motivated by the
facts that the identified particle spectra for nitrogen cover only
a small range in $z$, pions dominate "all charged" data and the
negative pion spectra are practically indistinguishable from the
positive ones. The comparison is shown in Fig. 3.

\begin{figure}[h]
\centerline{\epsfig{figure=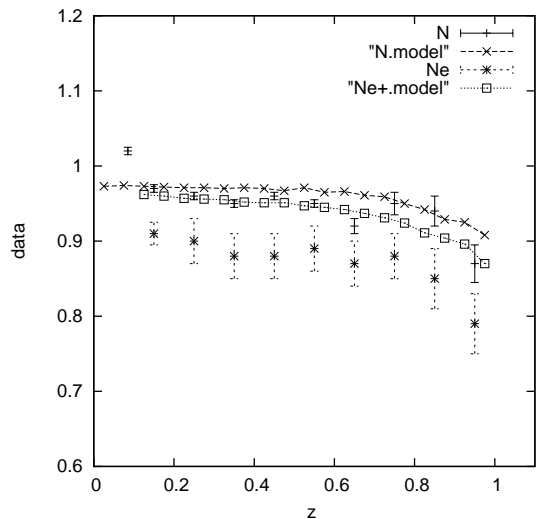, height =7cm}}
\caption{\label{fig3} {\small \sl The experimental ratio of the
charged hadrons $z$-spectra for nitrogen to deuterium \cite{H1}
and the positive pions for neon to deuterium \cite{H3} compared
with the model calculations.}}
\end{figure}
\par
We see clearly that the model predicts very little difference for
these two nuclei. This is understandable, as the difference in the
atomic number ($14$ vs $20$) corresponds to less than $15\%$ in
the value of nuclear radius. Thus the absorption effects measured
by the deviation of the ratio of spectra from $one$ should not
differ very much for the two nuclei.

\par
However, the data show a significant  difference. The lack of
fluctuations in the neon data indicates that the errors are
dominated by systematic effects. Still, these data suggest the
absorption effects twice as big as for the nitrogen. This
discrepancy is the main reason for the disagreement of our model
with new HERMES data.

\section{Conclusions}

~~~~~We have investigated
 the electroproduction of hadrons inside the nuclei using the PYTHIA
 event generator.
 The results from the recent HERMES experiment \cite {H3}
  are compared with the simple absorption model used earlier to
  describe the older
 data from the same experiment \cite{H1,H2}. We have found a surprising discrepancy.
 Its origin can be traced back to the unexpectedly large difference between the data
 for the nitrogen and neon nuclei. This difference seems to contradict any
 simple geometrical absorption picture. Thus any definite statements about the (dis)agreement
 of models with these data should be postponed until this difference is cleared
 out.\\

~~~~~ We thank Andrzej Bia{\l}as and Andrzej Kota{\'n}ski for
reading the manuscript and for helpful remarks. This work was
partially supported by the research grant 1 P03B 045 29
(2005-2008). One of us (RW) is also grateful for a partial
support  by the Marie Curie Actions Transfer of Knowledge project
COCOS (contract MTKD-CT-2004-517186).


\begin{thebibliography}{99}
\bibitem {FW} K. Fia{\l}kowski, R. Wit, hep-ph/0702058v2 (2007), to be published in Eur. Phys. J. A (2007).

\bibitem {H1} A. Airapetian {\sl et al.}, The HERMES Collaboration, {\it Eur. Phys. J.} {\bf C20}, 479 (2001);
A. Airapetian {\sl et al.}, The HERMES Collaboration, {\it Phys. Lett.} {\bf B577}, 37 (2003).
\bibitem {H2} A. Airapetian {\sl et al.}, The HERMES Collaboration, {\it Phys. Rev. Lett.} {\bf 96}, 162301 (2006).
\bibitem{PYTHIA}T. Sj\"ostrand {\sl et al.}, {\it Comp. Phys. Comm.}
 {\bf 135}, 238 (2001).
\bibitem {H3} A. Airapetian {\sl et al.}, The HERMES Collaboration, hep-ex/0704.3270v1 (2007).
\end{thebibliography}
\end{document}